\documentclass[12pt]{article}
\usepackage{amsmath,amsfonts,amssymb,amsthm,amstext,amscd,eucal}
\usepackage[all]{xy}

\makeatletter \@addtoreset{equation}{section}

\makeatletter\renewcommand\section{\@startsection {section}{1}{\z@}%
                                   {-3.5ex \@plus -1ex \@minus -.2ex}%nn
                                   {2.3ex \@plus.2ex}%
                                   {\normalfont\large\bfseries}}
\renewcommand\subsection{\@startsection{subsection}{2}{\z@}%
                                     {-3.25ex\@plus -1ex \@minus -.2ex}%
                                     {1.5ex \@plus .2ex}%
                                     {\normalfont\bfseries}}

\newcommand{\ie}{{\em i.e. }}

\newcommand{\cf}{\emph{cf.~}}

\newcommand{\be}{\begin{equation}}
\newcommand{\ee}{\end{equation}}
\newcommand{\bea}{\begin{eqnarray}}
\newcommand{\eea}{\end{eqnarray}}

\newcommand{\bse}{\begin{subequations}}
\newcommand{\ese}{\end{subequations}}
\newcommand{\bi}{\begin{itemize}}
\newcommand{\ei}{\end{itemize}}

\newcommand{\mpl}{M_{\rm pl}}

\def\Tr{  \mbox{Tr}   }

\begin{document}
\begin{titlepage}

\begin{flushright}\vspace{-3cm}
{\small
{\tt arXiv:1101.0048} \\
UUITP-43/10,\\
IPM/P-2011/001 \\
\today }\end{flushright}
\vspace{-.5cm}

\begin{center}
\centerline{{\Large{\bf{Gauged M-flation, its UV sensitivity and Spectator Species}}}} \vspace{4mm}

{\large{{\bf A.~Ashoorioon\footnote{e-mail: amjad.ashoorioon@fysast.uu.se}$^{,a}$,
M.M. Sheikh-Jabbari\footnote{e-mail:
jabbari@theory.ipm.ac.ir}$^{,b}$}}}
\\

%\vspace{5mm}

\bigskip\medskip
\begin{center}
%\centerline
{$^a$ \it Institutionen f\"{o}r fysik och astronomi
Uppsala Universitet,\\ Box 803, SE-751 08 Uppsala, Sweden}\\
\smallskip

%\centerline
{$^b$ \it School of Physics, Institute for Research in Fundamental
Sciences (IPM),\\ P.O.Box 19395-5531, Tehran, Iran}\\
%\smallskip
\end{center}
%\vfil
\vspace{5mm}

\end{center}
\setcounter{footnote}{0}

\date{\today}

\begin{abstract}

In this paper we study gauged M-flation, an inflationary model in which inflation is driven by three $N\times N$ scalar field matrices in the adjoint representation of $U(N)$ gauge group. We focus our study on the gauged M-flation model which could be derived from the dynamics of a stack of D3-branes in appropriate background flux. The background inflationary dynamics is unaltered compared to the ungauged case of \cite{Ashoorioon:2009wa}, while
the spectrum of ``spectator species'', the isocurvature modes, differs from the ungauged case. Presence of a large number of spectators, although irrelevant to the slow-roll inflationary dynamics, has been argued to lower the effective UV cutoff $\Lambda$ of the theory from the Planck mass $\mpl$, putting into question the main advantage of M-flation in not having super-Planckian field values and unnaturally small couplings. Through a careful analysis of the spectrum of the spectators we argue that, contrary to what happens in N-flation models, M-flation is still UV safe with the modified (reduced) effective UV cutoff $\Lambda$, which we show to be of order $(0.5-1)\times 10^{-1}\mpl$.
Moreover, we argue that the string scale in our gauged M-flation model is larger than $\Lambda$ by a factor of $10$ and hence one can also neglect stringy effects. We also comment on the stability of classical inflationary paths in the gauged M-flation.

\end{abstract}

\end{titlepage}
\renewcommand{\baselinestretch}{1.1}

%%%%%%%%%%%%%%%%%%%%%%%%%%%%%%%%%%%%%%%%%%%%%%%%%%%%%%%%%%%%%%%%%%%%%%5
\section{Introduction}

The idea of inflation in one of its simplest realizations involves a massive scalar field with potential $V=\frac{1}{2}m^2 \phi^2$ whose mass has to be hierarchically smaller than Planck mass, \ie $m\simeq 6\times 10^{-6}~\mpl$, where $\mpl\equiv (8\pi G_N)^{-1/2}= 2.43 \times 10^{18}$ GeV is the reduced Planck mass. The amount of scalar field displacement in the field space needed to produced the required $60$ e-folds of inflation in such model is much larger than Planck mass, explicitly $\Delta \phi \simeq 14~\mpl$. Moreover, recalling that the scalar field mass $m$ is quadratically sensitive to the UV cutoff, one should explain the hierarchy between $m$ and
$\mpl$. The flatness of potential, \ie $m\ll \mpl$, and super-Planckian field excursions $\Delta\phi\gtrsim \mpl$ poses both theoretical and model building challenges to inflationary scenarios and models.
For example, embedding of such a model in supergravity runs into various difficulties, since one has to guarantee the flatness of the potential on the scales which are beyond the limit of validity of the theory \cite{Lyth:1996im}.  The super-Planckian field excursions are also troublesome recalling that $\mpl$ is the ultimate UV cutoff in a theory coupled to (Einstein) gravity and hence one \emph{may} worry  about uncontrollable quantum corrections to the potential.\footnote{Another point of view is that having a super-Planckain field values in itself does not mean that the quantum gravity effects are important. From this perspective, the super-Planckian energy density is responsible for triggering the quantum gravity effects and thus in a chaotic slow-roll inflation, with $H\sim 10^{-5}\mpl$ and energy density way below $\mpl^4$, one may not worry about quantum gravity effect \cite{Linde:2004kg}.}
In supergravity or string theory motivated models  one usually finds the size of the region in which inflation can happen to be around $\mpl$ and it is therefore  not possible to motivate  the large field models like the one with quadratic potential, e.g. see \cite{SUGRA-inflation}. {It is, however, noteworthy that despite these model building problems, large field models are of phenomenological interest in anticipation of possible detection of B-mode polarization in the CMB, since they generally lead to sizeable primordial gravity waves.}

Inspired by the idea of assisted inflation \cite{Liddle:1998jc}, one way out of the super-Planckian field excursion problem was examined in \cite{Kanti:1999vt}, where $N$ scalar fields with polynomial chaotic-type  potential cooperate to increase the Hubble friction and induce an inflation with enough number of e-folding. For example, for an assisted inflation model with $N$ equal-mass scalar fields with quadratic potential, $V=\frac{1}{2}m^2 \sum_{i=1}^N\phi_i^2$, the N-flation \cite{Dimopoulos:2005ac}, one finds that $\Delta\phi_i\simeq 14\mpl/\sqrt{N}$. Hence, by sufficiently increasing the number of scalar fields one can lower the amount of displacement of each field to below $\mpl$. Nonetheless, even in this approach, there is no justification for the hierarchy between $m$ and $\mpl$. The situation is different for $N$ identical decoupled scalar fields with polynomial potentials other than the quadratic one, in which the process of making the kinetic term canonical, scal
 es the related couplings by negative powers of $N$. For cubic and quartic potentials, this will reduce the couplings by a fact
or of $\sqrt{N}$ and $N$, respectively, \ie the inflationary trajectory for an N-flation model with the potential  $V(\phi_i)=\sum_{i=1}^N \frac{\hat\lambda}{4}\hat\phi_i^4-\frac{2\hat\kappa}{3}\hat\phi_i^3$ effectively behaves like a single field theory $\phi$ with the potential $V(\phi)=\frac{\lambda}{4}\phi^4-\frac{2\kappa}{3}\phi^3$ where
\be\label{dressed-factors-N-flation}
\phi=\frac{\hat\phi}{\sqrt{N}}\,,\qquad \kappa=\frac{\hat\kappa}{\sqrt{N}}\,,\qquad \lambda=\frac{\hat\lambda}{N}\,.
\ee
One can thus justify the smallness of the effective couplings for the inflaton by increasing the number of scalar fields arbitrarily. For the case of quartic potential the value observed curvature perturbations, together with demanding 60 e-folds, gives $\lambda\sim 10^{-14}$ and field excursions $\Delta\phi\sim 10\mpl$ {\cite{Linde:1983gd}}. Therefore, requirement of having natural physical couplings, \ie $\hat\lambda\sim {\cal O}(1)$, is met if $N\sim 10^{14}$. The value of physical field excursions is then $\Delta\hat\phi_i\sim 10^{-6} \mpl$.

Recently, motivated by the dynamics of $N$ D3-branes subject to a proper RR six-form in a pp-wave background, an inflationary model was introduced  \cite{Ashoorioon:2009wa,Ashoorioon:2009sr}. Due to the matrix nature of the inflatons, we dubbed this model as  Matrix Inflation, or M-flation for brevity.  In this model, three scalar fields corresponding to three  dimensions perpendicular to the D3-branes play the role of the inflaton. In some specific representation for the matrices,  the SU(2) sector, the dynamics of the system could be mapped to a single scalar field with a fourth order polynomial potential whose cubic and quartic couplings are lowered, respectively, by factors of $N^{-3/2}$ and $N^{-3}$ for large $N$ \cite{Ashoorioon:2009wa}.  The model is nonetheless not an assisted model, as  in the assisted model all the scalar fields move in a concerted way to increase the Hubble friction and realize inflation. Thus, one should not expect the attractor behavior observed
in assisted models \cite{Liddle:1998jc,Postma:2010wd}. The model has this additional virtue of attaining the smallness of the couplings in generic chaotic inflation by a less number of degrees of freedom during inflation; this is achieved by $N\sim 10^{5}$ for M-flation compared to $N\sim 10^{14}$ of N-flation.\footnote{Note that in M-flation, where we deal with $N\times N$ matrices, the number of degrees of freedom grows like $N^2$, which is still lower by four orders of magnitude  than the N-flation number of degrees of freedom.}

The above advantages and successful features of N-flation, assisted inflation or M-flation is challenged by {the claim of \cite{Dvali:2007hz}}, stating that in the presence of $N_s$ ``light'' species the universal gravitational cutoff is not $\mpl$; it is $\Lambda$,
\be\label{species-cutoff}
\Lambda=\frac{\mpl}{\sqrt{N_s}}\,.
\ee
One may first make sure that the arguments of \cite{Dvali:2007hz,Dvali:2008jb} which has been mainly based on black hole physics considerations is also applicable to cosmological FRW setups. This we will argue for in section \ref{section-3}. In this case, nonetheless, as we will discuss  only the modes with (effective) masses below the Hubble parameter contribute to the number of species $N_s$. For the case of $m^2\phi_i^2$ N-flation model discussed above, where successful inflation implies $m\sim 10^{-6}\mpl$, with $H\sim 10^{-5}\mpl$, all of the $N$ fields contribute to the species counting and hence $N_s\simeq N$. (A similar result is true for other chaotic N-flationary models discussed above.) This means that physical field excursions and the effective gravitational cutoff are lowered in the same way, by the factor $1/\sqrt{N}$. As such, $\Delta\hat\phi\sim 10\Lambda$. That is, although the naturalness problem for  ${m/\Lambda}$ --taking $N\sim 10^{12}$-- or for the coupl
 ing $
 \hat\lambda$ --taking $N\sim 10^{14}$-- is solved, ``larger-than-cutoff field excursion problem'' is resurfaced again \cite{Postma:2010wd}.

In this paper we revisit M-flation in view of the above lowered UV cutoff $\Lambda$. Some preliminary analysis in this direction has already been made in \cite{Postma:2010wd}. As we will show, unlike N-flation, M-flation is safe from the above mentioned UV problem. This is due to the specific feature of isocurvature modes of M-flation: the mass spectrum of isocurvature modes (which following \cite{Postma:2010wd} will be called ``spectator modes'')
of M-flation contains
a variety of masses, with masses which are parametrically both above and below the Hubble parameter. Therefore, $N_s$ does not scale like $N^2$. As we will show, for the value of parameters  fixed by demanding having a successful M-flation model, $N_s$ is obtained to be of order 100-1000, while ${N}\sim 10^5$. This will save M-flation from the reappearance of the ``{larger-than-cutoff}'' problem.

The outline of this work is as follows. In section \ref{The-setup}, we review the setup of M-flation. In this work, however, we consider a specific ``gauged M-flation'' model, which is more closely related to D-brane dynamics. We then compute the spectrum of spectators of the gauge M-flation model. In section \ref{section-3}, we review the arguments of \cite{Dvali:2007hz} resulting in the UV cutoff modification \eqref{species-cutoff} and extend those arguments to the cosmological FRW universe. In section \ref{M-flation-Lambda-section}, we confront gauged M-flation with the modified (lowered)  gravitational UV cutoff and show that it is UV safe. The final section contains our concluding remarks and discussions. Some details of the arithmetics of the specific gauged M-flation model considered in this paper in a couple of inflationary regions has been gathered in the Appendix.

\section{Gauged M-flation, the setup}\label{The-setup}

M-flation, or Matrix inflation is the  model in which inflation is
driven by three $N\times N$ hermitian matrices $\Phi_i$ ($i=1,2,3$)
as inflaton fields with a specific quartic potential
\cite{Ashoorioon:2009wa,Ashoorioon:2009sr}. The M-flation model of
\cite{Ashoorioon:2009wa} has a \emph{global} $U(N)$ symmetry and
$\Phi_i$ are in its adjoint representation. On the other hand, $N$
string theory D3-branes probing specific background geometry
provides a natural setting in which M-flation can be realized. In
the brane theory setting, however, this $U(N)$ is gauged. In this
work, motivated by the string theory picture,  we will  focus on the
``gauged M-flation model'' the
action for which is%
\be\label{action}%
 S=\int d^{4} x \sqrt{-g} \left(\frac{-M_{P}^{2}}{2} R - \frac{1}{4} \Tr(F_{\mu\nu}F^{\mu\nu})- \frac{1}{2}
\sum_{i} \Tr  \left( D_{\mu} \Phi_{i} D^{\mu} \Phi_{i}
\right) - V(\Phi_{i}, [ \Phi_{i}, \Phi_{j}] ) \right) \, , %
\ee %
where the signature of the metric is $(-,+,+,+)$, $D_\mu$ is the
gauge covariant derivative and $F_{\mu\nu}$ is the gauge field strength: %
\be%
D_\mu\Phi_i= \partial_\mu \Phi_i+i g_{YM}[A_\mu,\Phi_i]\,,\qquad
F_{\mu\nu}=\partial_\mu A_\nu-\partial_\nu
A_\mu+ig_{YM}[A_\mu,A_\nu]\,,
\ee%
and the  $\Tr$ is over $N\times N$ matrices. The potential
$V(\Phi_{i}, [ \Phi_{i}, \Phi_{j}])$ can be motivated from dynamics
of $N$ D3-branes subject to an RR six-form, whose strength is
parameterized by $\hat\kappa$ and has  two legs along the directions
transverse to the D3-branes, in a specific ten-dimensional IIB
supergravity background \cite{M5-brane}:\footnote{Strictly speaking,
in order to view \eqref{action}, with a nonvanishing  four
dimensional Planck mass, as the low energy effective theory of $N$
D3-branes in the background \eqref{sugra-background}, we need to
demand  the six dimensional transverse space to be compact. This
could be achieved if we considered a background geometry which
around $x^i\sim l_s$ behaves like \eqref{sugra-background} and at
large values of $x^i$ becomes a (Ricci) flat geometry, which is then
compactified on a $T^6$ or CY$_3$. This latter is similar to the
standard KKLMMT scenario \cite{KKLLMT} where the $AdS_5\times S^5$ throat is
completed into an $R^4\times CY_3$.}
\bea\label{sugra-background}%
ds^2 &=&-2dx^+dx^--\hat m^2 \sum_{i=1}^3 (x^i)^2 (dx^
+)^2+\sum_{I=1}^8d
x_I dx_I, \\
C_{+123ij}&=& \frac{2\hat\kappa}{3} \epsilon_{ijk} x^k\,.
\eea%
The matrices $\Phi_{i}$ are proportional to three out of six dimensions transverse to the D3-branes
\be%
\Phi_i\equiv \frac{{X_i}}{\sqrt{(2\pi)^3 g_s}\ l_s^2}\,,
\ee%
and the potential takes the form%
\be\label{The-Potential}%
V= \Tr  \left( - \frac{\lambda}{4}  [ \Phi_{i},
\Phi_{j}] [ \Phi_{i}, \Phi_{j}] +\frac{i \kappa}{3} \epsilon_{jkl}
[\Phi_{k}, \Phi_{l} ] \Phi_{j} +  \frac{m^{2}}{2}  \Phi_{i}^{2}
\right),%
\ee%
where $i=1,2,3$ and  we are hence dealing with $3N^2$ real scalar
fields. Here and below the summation over repeated $i,j$ indices is
assumed. $\lambda$ and $\kappa$ are related to the string coupling
and the strength of the Ramond-Ramond antisymmetric form and $m$ is
the same $\hat m$ that appears in the metric:
\be%
\lambda=8\pi g_s=2g^2_{YM}\ ,\qquad \kappa= \hat{\kappa} g_s\sqrt{8\pi g_s}\ ,\qquad m^2=\hat m^2 . %
\ee%
From the string theory perspective, we need to choose $\hat m^2$ and
$\hat \kappa$ such that \eqref{sugra-background} is a solution to
supergravity equation of motion with a constant
dilaton, \ie %
\be\label{susy-cond}%
\lambda m^2=4\kappa^2/9\,. %
\ee %
{In \cite{Ashoorioon:2009wa}, we relaxed the above relation between the parameters of
the potential and construct more general M-flation models by
treating $\lambda,\ \kappa$ and $m^2$ as independent parameters.} In this
work we will restrict ourselves to gauged M-flation models with
\eqref{susy-cond} relation between its parameters.

\subsection{Background inflationary trajectory}

The equations of motion for the scalar and vector fields is given by
\be\label{e.o.m}%
\begin{split}
& D_\mu D^\mu \Phi_i+\lambda[\Phi_j,[\Phi_i,\Phi_j]]-i\kappa \epsilon_{ijk}[\Phi_j,\Phi_k]-m^2\Phi_i=0\,,\cr
& D_\mu F^{\mu\nu}-ig_{YM}[\Phi_i,D^\nu\Phi_i]=0\,.
\end{split}
\ee
As discussed in \cite{Ashoorioon:2009wa}, for the ungauged case, one can consistently
restrict the classical dynamics to a sector in which we are
effectively dealing with a single scalar field $\hat \phi$. This
sector, which will be called the SU(2) sector,
is obtained for matrix configurations of the form%
\be \label{phiJ}%
\Phi_{i} =
\hat \phi(t) J_{i}\ , \quad \quad i=1,2,3,%
\ee%
where $J_{i}$ are the basis for the $N$ dimensional irreducible
representation of the $SU(2)$ algebra%
\be\label{J}%
 [ J_{i}, J_{j} ]=  i \, \epsilon_{ijk} J_{k} \ , \qquad \Tr (J_{i} \,
J_{j})= \frac{N}{12} ( N^{2}-1 ) \, \delta_{ij} \, .%
\ee%
Since both $\Phi_{i}$ and $J_{i}$ are hermitian, we conclude that
$\hat \phi$ is a real scalar field.

It is straightforward to show that the $[\Phi_i, D_\nu\Phi_i]$ term in the equation of motion of the gauge field for the ansatz \eqref{phiJ} is proportional to $[J_i,[J_i,A_\nu]]$ and hence vanishes for $A_\mu=0$. Therefore,
in the SU(2) sector one  can consistently turn off the
gauge fields $A_\mu$ in the background, \ie the classical
inflationary trajectory takes place in the scalar fields $\Phi_i$
sector.\footnote{As discussed in \cite{gaugeflation} it is possible
to construct inflationary models where certain combination of the
non-Abelian gauge fields play the role of effective inflaton. For
the latter, however, one should have a specific gauge field action.}

Plugging these into the action (\ref{action}) and adding the four-dimensional Einstein gravity, we obtain
\be S= \int
d^{4} x \sqrt {-g} \left[- \frac{M_{P}^{2}}{2} R+ \Tr J^{2}  \left( -
\frac{1}{2}  \partial_{\mu} \hat \phi  \partial^{\mu} \hat \phi
-\frac{\lambda}{2}  \hat \phi^{4}  + \frac{2 \kappa}{3} \hat
\phi^{3} - \frac{m^{2}}{2} \hat \phi^{2} \right) \right] \, , \ee
where $\Tr J^{2} = \sum_{i=1}^{3} \Tr (J_{i}^{2})   = N(N^{2}
-1)/4$.
Interestingly enough, this represents the action of  chaotic
inflationary models with a non-standard kinetic energy. Upon the
field redefinition%
\be \label{phi-scaling}%
 \hat \phi = \left(  \Tr
J^{2}   \right)^{-1/2} \phi = \left[ \frac{N}{4}(N^{2}-1)
\right]^{-1/2} \, \phi \, , %
\ee%
the kinetic energy for the new field $\phi$ takes the canonical
form, while the potential becomes
\be\label{Vphi}%
V_0(\phi)= \frac{\lambda_{eff}}{4} \phi^{4} -
\frac{2\kappa_{eff}}{3} \phi^{3} + \frac{m^{2}}{2} \phi^{2} \, , %
\ee%
where%
\be \label{lameff}%
\lambda_{eff} = \frac{2 \lambda}{\Tr J^{2}} = \frac{8 \lambda}{ N
(N^{2}-1)}  \ , \quad \kappa_{eff} = \frac{ \kappa}{\sqrt{\Tr J^{2}}
} = \frac{2
\kappa}{\sqrt{N(N^{2}-1)}}\,. %
\ee%
Depending on the choice of parameters, $\lambda,\ \kappa$ and $m^2$,
several inflationary scenarios could be realized in Matrix inflation
setup which was studied in \cite{Ashoorioon:2009wa}. In this work we
will focus on the ``symmetry breaking'' case where these parameters
are related as in \eqref{susy-cond}. Our aim here is to study the
effective excursions of the field with the species-reduced UV cutoff
\cite{Dvali:2007hz} and examine the UV stability of the M-flation
scenario.

\subsection{Spectrum of gauged M-flation ``spectator'' modes}

In the gauged M-flation we start with three $N^2$ scalar fields
$\Phi_i$ and $4N^2$ fields $A_\mu$. Reducing to the SU(2)  sector
\eqref{phiJ}, we have turned on only one combination these fields at
the background level. Out of the remaining $7N^2-1$ fields,
recalling the gauge symmetry of the action, we expect $2N^2$ of the
gauge fields and $3N^2-1$ of the scalars to be physical; the $2N^2$
of the gauge field degrees of freedom is removed by the equations of
motion and gauge invariance. As we will show momentarily, although
the total count of $5N^2-1$ isocurvature modes remains intact, due
to the (spontaneous) symmetry breaking induced by the potential
$V(\Phi_i)$, we will have $3N^2-1$ vector field degrees of freedom
and $2N^2$ scalar modes. These $5N^2-1$ modes, however, can be
excited quantum mechanically. As was discussed in
\cite{Ashoorioon:2009wa}  backreaction of these isocurvature modes on the inflationary background,
at least during slow-roll  period, is very small. Moreover, these isocurvature modes do not couple to the quantum fluctuations of the effective inflaton field $\phi$. We will hence,
following \cite{Postma:2010wd}, call these modes  as \emph{spectators}.
The arguments of \cite{Ashoorioon:2009wa} are made for the ungauged
case and, as we will show below, it is straightforward to extend those
to the gauged case. We start our analysis by working out the \emph{spectators}
mass spectrum.

\paragraph{Spectrum of scalar spectators.}

The analysis of the spectrum for these modes is  the same as the
ungauged case of \cite{Ashoorioon:2009wa}, except for the fact that
``zero-modes'' are not physical in the gauged case (see below). To
compute the spectrum of scalar fluctuations
$\Psi_i$, defined as %
\be\label{Phi-expand}%
\Phi_i=\hat \phi J_i+\Psi_i\,, %
\ee%
we expand the action to second order in $\Psi$ while turning off the gauge fields,
yielding%
\be\label{Psi-action}%
{\cal L}_\Psi^{(2)}=  -\frac12\Tr(\partial_\mu\Psi_i)^2+\Tr\left[\frac\lambda2\hat\phi^2(\epsilon_{ijk}[J_j,\Psi_k])^2
+i(\frac{\lambda}{2}\hat\phi^2-\kappa\hat\phi)\epsilon_{ijk}[J_i,\Psi_j]\Psi_k -\frac{m^2}{2}\Psi_i^2\right]\,.%
\ee%
The above is ``diagonalized''  for $\Psi_i$ satisfying
\be%
i\epsilon_{ijk}[J_j,\Psi_k]=\omega\Psi_i\,.
\ee%
 The solutions of the above are spin one (vector) representations of SU(2). The details may be found in \cite{DSV}, here we only quote the result:
\begin{itemize}
\item \emph{zero modes:} $\omega=-1$
\be
i\epsilon_{ijk}[J_j,\Psi_k]=-\Psi_i\,.
\ee
There are $N^2-1$ of these modes. In the ungauged theory \cite{Ashoorioon:2009wa} these modes are physical, while in the gauged model they are not, as they are gauge degrees of freedom. To see the latter let us recall that under a global infinitesimal gauge transformation  $\Phi_i\to \Phi_i+ig[\Lambda, \Phi_i]$ where $\Lambda$ is a generic
$N\times N$ hermitian matrix. Therefore, a $\Psi_i$ in the zero mode is nothing but a gauge transformation over the background solution $\Phi_i=\hat\phi J_i$. We hence discard these modes as unphysical in our current analysis.

\item $\alpha_j$-modes: $\omega=-(j+2)$ and $0\leq j\leq N-2$. Degeneracy of each $\alpha_j$ mode is $2j+1$. There is therefore, $(N-1)^2$ of these modes. Recalling \eqref{Psi-action}, the mass of these modes are
\be%
M^2_{\alpha_j}= \frac12\lambda_{eff}\phi^2 (j+2)(j+3)-2\kappa_{eff}\phi (j+2)+m^2\,.
\ee%
\item $\beta_j$-modes: $\omega=j-1$ and $1\leq j\leq N$. Degeneracy of each $\beta_j$-mode is $2j+1$ and hence there are $(N+1)^2-1$ of $\beta$-modes. Mass of $\beta_j$ mode is%
\be
M^2_{\beta_j}=\frac12 \lambda_{eff}\phi^2 (j-1)(j-2)+2\kappa_{eff}\phi (j-1)+m^2\,.
\ee%

\end{itemize}

\paragraph{Spectrum of gauge field spectators.}

To read the spectrum of the gauge fields we turn on $A_\mu$ and expand the action \eqref{action} to second order in $A_\mu$ while replacing for the value of $\Phi_i=\hat\phi J_i$. The second order gauge field action is then obtained to be%
\be%
 {\cal L}_{A_\mu}^{(2)}=  -\frac14\Tr(\partial_{[\mu}A_{\nu]})^2+
\frac12g^2_{YM}\hat\phi^2 \Tr([J_i,A_\mu][J_i,A_\mu])\,. %
\ee%
The mass spectrum can be read from the second term recalling that
$\Tr([J_i,A_\mu][J_i,A_\mu])=\Tr(A_\mu[J_i,[J_i,A_\mu]])$ and that
$[J_i,[J_i, X]]=\omega X$ eigenvalue problem has eigenvalues
$j(j+1)$ with $j=0,\cdots, N-1$ and degeneracy of each mode is
$2j+1$. Therefore, we are dealing with a system of massive vector
fields with masses
\be%
M^2_{A,j}=\frac{\lambda_{eff}}{4}\phi^2 j(j+1)\,.
\ee%
As we see $j=0$ mode is massless and corresponds to the $U(1)$ sector in the $U(N)$ matrices. Degeneracy of the vector field modes is hence $3(2j+1)$ for $j\geq 1$ modes and is two for $j=0$ mode.
(The factor of three has appeared due to the three polarizations of each massive vector field.)
We hence have $3N^2-1$ vector field modes.

In summary we have $(N-1)^2$  $\alpha$-modes, $N^2+2N$ $\beta$-modes
and $3N^2-1$ vector field modes, altogether $5N^2$ modes. The
$\alpha_{j=0}$ mode (which has degeneracy one) and mass
$M^2=6\lambda\hat\phi^2-4\kappa\hat\phi+m^2$ is the quantum
fluctuations of the SU(2) sector scalar effective inflaton field $\phi$ and is hence
the adiabatic mode while all the other $5N^2-1$ modes are
``isocurvature''.

\section{Effective gravity UV cutoff and number of species}\label{section-3}

Planck scale $\mpl$ in the Einstein gravity has two roles: 1)
Recalling that  $8\pi G_N=\mpl^{-2}$, $\mpl$ is the coupling of
\emph{classical} gravity and, 2) the energy scale above which
quantum gravity effects kick in; it is the ultimate UV cutoff for
the quantum field theories above which (quantum) gravity effects
cannot be ignored. There are, however, various perturbative and
non-perturbative arguments suggesting that the UV cutoff $\Lambda$
above which (quantum) gravity effects become important  is not
$\mpl$, and it can be a (much) lower scale
\cite{Dvali:2007hz,Dvali:2008jb,Dvali:2010vm}. Here we review some
of these arguments. The perturbative argument is as follows: suppose
we have $N_s$ quantum fields with masses $\Lambda$ coupled to
gravity. Each of these quantum fields will induce a factor
proportional to $\Lambda^2$ into the renormalizable Planck mass
\cite{Adler:1980bx,Adler:1980pg}. Modulo possible accidental
cancelations, this suggests that the effective contribution to the
Planck mass is proportional to $N_s\Lambda^2$. Of course this
perturbative argument only suggests that $N_s\Lambda^2\leq \mpl^2$.
As it is clear,  $N_s$ is the number of species whose mass is below
the cutoff scale $\Lambda$. This result receives backing from other
known physics, which we will review below. In the non-trivial
gravity backgrounds where there is an energy scale associated with
the background itself, the question of which degrees of freedom
contribute to the counting $N_s$ should be revisited. Below we
discuss two such cases: A (Schwarzchild) black hole and the FRW
cosmological background.

\paragraph{Black holes and the species cutoff.}
It is an established fact that a Schwarzchild black hole of mass
$M_{BH}$ is a thermodynamical system with the Bekenstein-Hawking
entropy \cite{Bekenstein:1973ur,Hawking:1974sw}
\be\label{BH-ent}%
S_{BH}=2\pi\mpl^2 A_h=\frac12 \left(\frac{M_{BH}}{\mpl}\right)^2\,,%
\ee%
where $A_h$ is the horizon area, and at Hawking temperature
\be\label{T-BH}%
T_{BH}=\frac{\mpl^2}{M_{BH}}\,. %
\ee%

As in any thermodynamical system, unitarity is lost \emph{unless}
there is an underlying unitary statistical mechanical description
which leads to the thermodynamical system in the ``thermodynamic
limit''. Despite the partial progress for some special cases, a
general statistical mechanical description of Bekenstein-Hawking
entropy is still lacking. One such attempt is to understand the
black hole entropy as the entanglement entropy of a system
accounting for the ``microstates'' of the black hole. If this system
consists of $N_s$ number of species lighter than the cutoff scale
$\Lambda$ this entanglement entropy is \cite{Israel:1976ur}
\be\label{entanglement-ent}%
S_{\rm ent}=N_s\frac{\Lambda^2}{\mpl^2} A_h\mpl^2\,.%
\ee%
Assuming  $\Lambda\simeq \mpl$  leads
to species problem: even though the Bekenstein-Hawking entropy is
universal, the  entanglement entropy is not and depends on the
number of species. Eq.\eqref{species-cutoff} can serve as a
resolution to this problem.

The above argument is not limited to interpreting the
Bekenstein-Hawking entropy as the entanglement entropy and can be
argued for noting the thermodynamical nature of the black hole. To
see this more clearly let us recall that the Hawking radiation of a
black hole consists of particles which it can thermally produce, \ie
their masses are less than its Hawking temperature \eqref{T-BH}. On
the other hand the semi-classical treatment of black hole is only
valid for $T_{BH}\lesssim \Lambda$ and when its energy emission rate
$dE/dt \lesssim \mpl^2$. The energy emission rate is proportional to
$N_s T_{BH}^4 A_h$, where $N_s$ is the number of (relativistic)
particles which can be produced by the black hole. Putting these
together, we learn that $N_s\Lambda^2\lesssim \mpl^2$.

One may also argue for this bound in a different way: Let us suppose
that we have a system of $N_s$ quantum fields of mass $m_0$ (we are
assuming that $m_0$ is less than the eventual cutoff $\Lambda$) and
that these $N_s$ fields are labeled by e.g. a discrete $Z_{N_s}$
symmetry. Semiclassical description  for a black hole is available
if its Hawking temperature is at most of  order the smallest mass
state available, \ie ${T_{BH}\gtrsim m_0}$. Moreover, due to no-hair
theorem, the black hole state should be $Z_{N_s}$ invariant and
hence its lowest mass is $N_s m_0$. These again imply that
$m_0^2\lesssim \mpl^2/N_s$. It is worth noting that this
arguments are compatible with the physical expectation that the
life-time of a semiclassical black hole should not be less than
$\Lambda^{-1}$.

\paragraph{FRW backgrounds and the species cutoff.}

For the case of a black hole horizon, our argument was mainly based
on the fact that there is a natural energy scale associated with the
system, the Hawking temperature $T_{BH}$ and that the effective
cutoff $\Lambda$ must be less than this temperature. For the cases
where we have a cosmological horizon this natural scale should be
replaced with the Hubble parameter of the space $H$. In a classic
paper Gibbons and Hawking \cite{Gibbons-Hawking} have argued that
for the cosmological event horizons indeed one can still use the
``first law of black hole thermodynamics'' in the same way as used
for black hole event horizons, but with $T_{BH}$ replaced
$\kappa_G/(2\pi)$ where $\kappa_G$ is the surface gravity at the
cosmological event horizon, which is nothing but the Hubble
parameter $H$.

The above has a direct manifestation in the well-established cosmic
perturbation theory, e.g. see \cite{Mukhanov-Book}: the amplitude of
quantum fluctuations of ``light'' fields, fields whose mass are
small compared to the Hubble radius $H$, at the horizon crossing, is
equal to the value set by the ``thermal'' fluctuations, which is
nothing but the Gibbons-Hawking temperature. In other words in a
cosmological setting species which contribute to the cutoff are
number of the fields whose mass is less than the Hubble parameter.
This is what we are going to use next to compute the effective
cutoff for the M-flation.

\section{Gauged M-flation and the species gravitational UV cutoff}\label{M-flation-Lambda-section}

Having worked out the spectrum of the gauged M-flation spectators we
are now ready to calculate the number of species $N_s$  contributing
to the effective cutoff $\Lambda$: \be N_s=\textrm{number of
spectators with mass less than}\ H\,,
\ee%
where $H$ is given by the Friedmann equation
\be
3H^2\mpl^2=\frac{\lambda_{eff}}{4}\mu^4 x^2(x-1)^2\,.
\ee%
In the above we have considered the ``symmetry breaking'' M-flation
model (see the Appendix for a more detailed inflationary analysis of
this case),
$$
\phi=\mu x\,,\qquad \mu^2=\frac{2m^2}{\lambda_{eff}}\,.
$$
Having a successful inflationary model implies that $x$ is a
parameter of order one during inflation, $\mu$ is of order $25-35$
$\mpl$ and $\lambda_{eff}\sim 5\times 10^{-14}$
\cite{Ashoorioon:2009wa}, see also the Appendix here.
In terms of $x$ and $\mu$ parameters the masses
are
\be\label{spectrum-mu}%
\begin{split}
M^2_{\alpha_j} &=\frac{\lambda_{eff} \mu^2}{2}\left[x^2(j+2)(j+3)-3x (j+2)+1\right]\,,\qquad 0\leq j\leq N-2\cr
M^2_{\beta_j} &=\frac{\lambda_{eff }\mu^2}{2}\left[x^2(j-1)(j-2)+ 3x (j-2)+1\right]\,,\qquad 1\leq j\leq N\cr
M^2_{A,j} &=\frac{\lambda_{eff} \mu^2}{4} x^2 j(j+1)\,,\qquad 0\leq j\leq N-1\,,
\end{split}
\ee%
The above mass spectra are increasing as we increase $j$; for large $j$
they grow like $j^2$. Noting that $H^2\gg \lambda_{eff}\mu^2$, the
heaviest mode which contribute to the
species count is then given by%
\be\label{j-max}%
j_{max}^2\simeq \frac{(x-1)^2}{6} \left(\frac{\mu}{\mpl}\right)^2%
\ee%
for $\alpha$ and $\beta$ modes, and twice as much for the gauge field
modes. Therefore, considering the degeneracy of state for a given
$j$, $N_s$ is given by
\be%
N_s \simeq(2+ 3\cdot 2) j_{max}^2=
\frac{4(x-1)^2}{3}\left(\frac{\mu}{\mpl}\right)^2\,. %
\ee%
In the above the factor of 2 is for $\alpha$ and $\beta$ modes and
$3\cdot {2}$ is for the gauge fields; 3 for the polarization of
massive gauge fields and ${2}$ for the extra factor of $2$ in the
spectrum of gauge fields compared to $\alpha,\ \beta$ modes
(\emph{cf} \eqref{spectrum-mu}).  The effective UV cutoff is then%
\be\label{UV-cutoff-M-flation}%
\frac{\Lambda^2}{\mpl^2}=\frac 1{N_s}=
\frac{3}{4(x-1)^2}\left(\frac{\mpl}{\mu}\right)^2\,,\quad \textrm{or} \quad \Lambda=\frac{2(x-1)}{\sqrt{3}}\ \frac{\mu}{N_s}\,.%
\ee%

We note that $N_s$ and hence $\Lambda$ only depend
on $\mu$ and not the other parameter of the model $\lambda_{eff}$,
whereas size of the matrices $N$ was fixed on the requirement of
having an order one $\lambda$ parameter. For a successful
``natural'' inflation (see the Appendix) $N\sim 5\times 10^4$ while
$N_s\sim (2-8)\times 10^2$. As we see $N_s$ is not only different
than $N^2$ as one would have naively thought, but also $N_s\ll N$.
For our model, as we see, it happens that numerically  $N_s\sim \sqrt{N}$ and $\Lambda\simeq 0.05 \mpl$.

It is instructive to compare the scale of energy density during
inflation with the cutoff scale, \ie $\rho/\Lambda^4$ ratio, where
$\rho$ is the energy density driving  inflation. One can show
that%
\be%
\frac{\rho}{\Lambda^4}=\lambda_{eff} N_s^4\
\frac{9 x^2}{64 (x-1)^2}\,.
\ee%
For the parameters of our model this ratio is $< 10^{-3}$. We would
like to stress that, although it happens for our model, the energy
density of the background $\phi$ field during inflation need not be
less than $\Lambda^4$. This is due to the fact that, as can be seen
from our discussions of the previous section, the suppression of the
effective cutoff with respect to $\mpl$ is only relevant to the
quantum fluctuations and not the background classical fields.

As discussed the M-flation is motivated by or derived from dynamics
of D-branes in string theory. The D-branes are, however, described
by Born-Infeld and our M-flation action is obtained from expansion
of the Born-Infeld in the the leading order in string scale $m_s$.
In using M-flation, one should then make sure that i) keeping the
first order terms in the Born-Infeld is a valid expansion and ii)
the stringy effects should not become important below the cutoff
scale $\Lambda$, \ie $m_s\gtrsim \Lambda$. One would also
physically expect $m_s\leq \mpl$. The ratio of the first two terms
in the expansion of Born-Infeld action for $N$ branes is given by
\be\label{ratio-stringy} \delta\equiv
\frac{\Tr([\Phi_i,\Phi_j]^4)}{m_s^2\ \Tr([\Phi_i,\Phi_j]^2)}\sim
(\frac{\hat\phi}{m_s})^2\ \frac{\Tr J^4}{\Tr{J^2}}\sim
\frac{\phi^2}{4N m_s^2}\,.
\ee%
In computing the above \eqref{phiJ}, \eqref{J} and
\eqref{phi-scaling}  and,  $\Tr J^n\sim N(N/2)^{n}$ approximation
have been used. Moreover, in the above estimate for $\delta$ in
``$\sim$'' we are missing factors of order $10^{-1}$. A good
estimate for $\delta$ is obtained by replacing $\phi$ by the value
of the field at its minimum $\mu$. This gives a lower bound
approximation for the $\phi>\mu$ inflationary model and an upper
bound for $\phi<\mu$ hilltop inflation case. Demanding $\delta
\lesssim 1$ yields \be {m_s}\gtrsim \frac{\mu}{\sqrt{N}}\,\quad
\textrm{or}\quad m_s\gtrsim \Lambda\ \frac{N_s}{\sqrt{N}}\,.
\ee%
For our case $\frac{N_s}{\sqrt{N}}\sim 1$ and hence $m_s\gtrsim \Lambda$. Therefore, one can safely ignore the stringy corrections and the Born-Infeld corrections to the M-flation action once the energy  remains below the cutoff $\Lambda$. With the numeric values for $N$ and $N_s$, $\Lambda=5\times 10^{-2}\mpl,\ m_s\sim 10^{-1}\mpl$ is a reasonable range.

\section{Concluding remarks}\label{discussion-section}

We demonstrated that in the gauged M-flation, and in its SU(2)
sector, the physical field excursion remains below the species UV
cutoff $\Lambda$. This is rooted in the fact that species which
contribute to the counting are the ones that are lighter than the
Hubble parameter $H$, which plays a role similar to the black hole
temperature in de-Sitter space. In gauged M-flation model, we find
the modified UV cutoff $\Lambda$ to be few percent of $\mpl$ due to
hierarchical nature of isocurvature modes mass spectra. Although we
focused our analysis on the gauged M-flation case, from the results
of \cite{Ashoorioon:2009wa} and our discussions of section 4, it is
obvious that a similar conclusion could be drawn for the ungauged
case too.

The situation is, however, different in chaotic assisted models
\cite{Kanti:1999vt} or N-flation \cite{Dimopoulos:2005ac} in which
all the isocurvature modes  have masses smaller than the Hubble
parameter. In this regard M-flation is more successful than these
models, as it is UV safe. The problem of excursion beyond the cutoff
remains an open question in those models. Of course, some assisted
scenarios like multiple M5-brane inflation
\cite{Becker:2005sg,Ashoorioon:2006wc} may survive, as in those models
the physical field excursions are scaled by $N^{-3/2}$. Even though
the background classical energy density need not be smaller than
$\Lambda^4$, the gauged M-flation energy density happens to respect
this bound. As discussed the effective mass parameter in the SU(2) sector \eqref{Vphi} and the original M-flation action \eqref{action} are the same and there is no $N$ scaling for the mass parameter. To have a successful inflation $m\sim 10^{-6}\mpl$ and hence $m/\Lambda \sim 10^{-4}$. That is, the hierarchy of the mass parameter is reduced (improved) by two orders of magnitude compared to simple $m^2\phi^2$ case.

We would also like to comment on the stability of the classical
inflationary trajectory in gauged M-flation with the symmetry
breaking potential, {with respect to the quantum production of $\Psi_i$ modes}. A similar analysis for the case of ungauged
$\lambda\Phi^4$ M-flation has been carried out in
\cite{Ashoorioon:2009wa}, showing that backreaction of the large
number of spectator fields during slow-roll inflation will always remain small. Following the
line of arguments in section 6 of \cite{Ashoorioon:2009wa}, one can
show that a similar result holds for the symmetry breaking case, in
both gauged and ungauged cases. In this respect, the
situation in the gauged case is better, as there are no ``zero
modes'' and the zero modes are replaced with more massive states in
the vector field fluctuations (the zero modes are replaced
by the ``longitudinal'' modes of massive vector fields). Being more
massive, it is harder to excite these modes and hence their
backreaction is also reduced. The species cutoff considerations
actually help with the above {stability} argument, since the modes
which are really excited are even less, modes with $j>j_{max}$ (\cf
\eqref{j-max}) will not enter into the backreaction analysis.

A large number of ``light species'' of isocurvature modes, besides the reduction of the effective gravitational cutoff discussed above, may give rise to the problem of ``dominance of quantum fluctuations'' of the isocurvature modes, effectively pushing the theory to the etenral inflation phase, rather than slow-roll inflation governed by the effective inflaton field. As was pointed out in \cite {Ahmad:2008eu},  in N-flation if the number of the fields goes beyond $N_{c}=\mpl^2/\bar{m}^2$, the slow-roll phase of inflation disappears. $\bar{m}^2$ is the r.m.s. of the masses of the fields that play role during inflation. As a result ordinary N-flation models suffer from this problem. In M-flation scenario, however, the hierarchical nature of masses in the directions orthogonal to the SU(2) sector saves the theory from this potential problem too. To see this let us
note that eternal phase of inflation starts when the total classical displacement of the inflaton(s), $\delta\phi_{\rm CL}$, becomes equal to its (their) quantum fluctuations, $\delta\phi_{\rm QM}$, \cite{Linde:1986fc}. The total amount of quantum fluctuations in M-flation is
\be
\delta\phi_{\rm QM}=\sum_{i=1} \frac{H}{2\pi}\,,
\ee
where $i$ runs over all the modes that are lighter than the Hubble parameter, $N_s$ for M-flation. On the other hand, the total amount of classical displacements is given by
\be
\delta\phi_{\rm CL}=\left|\frac{V'_0(\phi)}{3H^2}\right|=\mpl^2 \left|\frac{V'_0(\phi)}{V_0(\phi)}\right|
\ee
where $V_0(\phi)$ is the collective potential which results from all the fields, eq.\eqref{Vphi}. For different regions of our M-flation scenario, one can show that
$\delta\phi_{QM}$ is subdominant  to the $\delta\phi_{\rm CL}$. For example, for $\phi>\mu$, $\delta\phi_{\rm QM}\simeq 0.003 \mpl$ and
$\delta\phi_{CL.}\simeq 0.16 \mpl$ around $\phi_{\rm ini}\simeq 43.5 \mpl$. Thus the problem of dominance of quantum fluctuations in N-flation will never occur in the case of M-flation, even if the number of D3-branes is increased.

It is also worthwhile to note that the gauged M-flation in the region $\phi>\mu/2$ is a local attractor for any perturbation about the SU(2) sector trajectory. This could be seen form the fact that the mass squared of $\alpha_j$,  $\beta_j$ and all the gauge modes, except for the $j=0$ gauge mode\footnote{$j=0$ gauge mode is massless and in the spectrum it corresponds to the U(1) sector of the gauge
fields. Being in the U(1) sector, this massless mode does not couple to the scalar fields $\Phi_i$.}, is positive around the $\Psi_i=0$ trajectory, if gauged M-flation happens in the region $\mu/2<\phi$. \footnote{For M-flation in the region $\mu/2<\phi<\mu$, the lightest spectator is $\alpha_0$. $M^2_{\alpha_0}$ is negative during about the first 34  e-folds of inflation, while all the other spectators have positive mass squared for the whole inflationary period.  On the other hand, we note that the $\alpha_0$ mode is nothing but the fluctuation of the effective inflaton $\phi$ itself, and the negative mass squared for some period  during initial phase of inflation is a generic feature of any hilltop inflation. Existence of this modes is hence not inducing any instability in the reduction to the SU(2) sector. {On the other hand, M-flation in the region $0<\phi<\mu/2$, besides the $ M^2_{\alpha_0}$, the first $\sim 93$ $\alpha-$modes become tachyonic, at least for a while, du
 ring the $60$ e-folds of inflation. Thus, in this M-flationary region, one expects perturbations around the SU(2) sector in these directions to grow, causing instability in the SU(2) sector inflationary trajectory.}} This guarantees that $\Psi_i=0$ is at least a local attractor and departures from the SU(2) sector will ultimately vanish. The rate of approaching the attractor is set by the ratio of $\Psi_i$-mass over the Hubble parameter and therefore deviations in the directions of ``small group'' of light modes will take $H/m$ e-folds to reach the SU(2) attractor. However, the situation is still much better than the assisted N-flation in which ``all'' the orthogonal modes are lighter than the Hubble parameter. If the UV cutoff is at the Planck mass, the perturbations of the $\Psi-$modes that are heavier than the Hubble parameter will reach the SU(2) sector in much less than an e-fold.

Finally we comment that, although  the background \eqref{sugra-background} is a solution to type IIB supergravity and one can consistently study dynamics of $N$ D3-branes in this background, the directions transverse to branes are noncompact and hence the effective four dimensional Newton constant goes to zero. In order to have a precise string theory derivation of \eqref{action}, we  need to complete the background \eqref{sugra-background} by viewing it as a part of the geometry which has six compact directions of finite volume. This latter is conceptually similar to usual mobile brane inflationary models where the brane is moving in an AdS throat while the geometry away from the throat is
$R^4\times CY3$. Moreover, given this ``completed" geometry one should make sure that the backreaction from the stack of $N$ D3-branes would remain negligible. We will postpone the study of these issues and a precise string theory realization of M-flation  to a future publication.

\section*{Acknowledgements}

A.A. was supported by the G\"{o}ran Gustafsson Foundation.

\appendix
\section{Symmetry-breaking inflation}\label{Appendix}
%$\lambda m^2=4\kappa^2/9$

In this case the potential takes the form of a symmetry-breaking potential which has two global minima at $\phi=0$ and $\phi=\mu$
\be\label{Vsusy}%
V_0=  \frac{\lambda_{eff}}{4} \phi^{2}  \, (\phi- \mu)^{2},
\ee%
where $\mu \equiv \sqrt 2 m/\sqrt{\lambda_{eff}}$. The minimum at $\phi=\mu$ corresponds to
supersymmetric vacuum when $N$ D3-branes blow up into a giant
D5-brane in the presence of background RR field. The minimum at $\phi=0$, on
the other hand, corresponds to the trivial solution when matrices
become commutative. If we allow the field $\phi$ to take negative
values, then the potential is symmetric under $\phi \rightarrow
-\phi + \mu$.

Depending on the initial value of the inflaton field, $\phi_{i}$,
the inflationary can take place in three different regions:

{\bf (a)}~$ \phi_{i}> \mu$

Suppose inflation starts when $ \phi_{i}> \mu$. With $N_{e}=60, \delta_{H} \simeq 2.41 \times 10^{-5}$ and $n_{s}=0.96$, one obtains%
 \be \label{case1}%
 \phi_{i} \simeq 43.57
\mpl \,, \qquad \phi_{f} \simeq 27.07 \mpl \,,\quad  \quad \mu
\simeq 26 M_{P}\,.%
 \ee%
  and
\begin{equation}\label{phi>mu}
   \lambda_{eff}\simeq 4.91 \times 10^{-14}, \quad m\simeq 4.07\times 10^{-6} \mpl, \quad \kappa_{eff}\simeq 9.57 \times 10^{-13} \mpl.
\end{equation}
Assuming that $\lambda=1$, in order to obtain the desire value for
$\lambda_{eff}$, one needs $N=54618$ D3-branes. The amount of
excursion in the physical field space, $\Delta\hat \phi$, is %
\be%
\frac{2(\phi_f-\phi_i)}{\sqrt{N(N^2-1)}}=2.58\times 10^{-6}
\mpl\,.%
\ee%
The Hubble parameter in the beginning of inflation, $H_i\equiv
H(\phi_i)=4.89\times 10^{-5}\mpl$. Maximum $j$ for which
$\alpha$-modes, $\beta$-modes and gauge modes are smaller than
$H_i^2$, respectively, are%
\bea%
j_{max}^{\alpha}=5,\quad j_{max}^{\beta}=7, \quad j_{max}^{g}=9\,.
\eea%
The total number of species that would contribute to the cutoff is
then $N_s=398$. The UV cutoff, $\lambda$, is found to be
\begin{equation}\label{Lambda-calculation-appendix}
\Lambda=5\times 10^{-2} \mpl\,,
\end{equation}
which is much larger than the field displacement. The ratio $\rho/\Lambda^4$ is $1.14\times 10^{-3}$.

{\bf (b)}~$ \mu/2<\phi_{i}< \mu$

To fit the observational constraints one obtains%
\be \label{case2}%
\phi_{i} \simeq 23.5 \mpl \,,  \qquad \phi_{f} \simeq 35.03 \mpl \,,
\qquad \mu \simeq 36 M_{P}\,.%
\ee%
and%
\begin{equation}\label{phi<mu}
\lambda_{eff}\simeq 7.18\times 10^{-14}\,, \qquad m\simeq 6.82\times
10^{-6} \mpl\,, \qquad \kappa_{eff} \simeq 1.94\times 10^{-12}
\mpl\,.
\end{equation}%
Repeating the same analysis for this branch, one realizes that $N=48103$ D3-branes are needed. The amount of physical excursions of the field is $2.185\times 10^{-6}~\mpl$. Number of species that contribute to the cutoff is $N_s=195$ and the corresponding UV cutoff is $\Lambda=7.16
\times 10^{-2}\mpl$. The ratio of energy density of the Universe during inflation to $\Lambda^4$ is $5.89\times 10^{-4}$.

{\bf (c)}~$ 0<\phi_{i}< \mu/2$

Due to symmetry $\phi \rightarrow -\phi + \mu$ this inflationary
region has the same properties as $ \mu/2 <\phi_{i}< \mu$ above and therefore the couplings have the same values as region {\bf (b)}. However the mass expressions for the isocurvature modes do not enjoy the symmetry and therefore the numerics are a little bit different from case {\bf (b)}. The total number of species contributing to the cutoff in this case is $N_s=794$ and the corresponding value for the UV cutoff is $3.55\times 10^{-2}\mpl$. The ratio of the energy density of the Universe to $\Lambda^4$ is $9.76\times 10^{-4}$.

\end{document}